\PassOptionsToPackage{table}{xcolor}
\documentclass[sigconf,nonacm, authorversion]{acmart}

\usepackage[T1]{fontenc}
\usepackage{subcaption}
\usepackage{booktabs}   
\usepackage{longtable}  % Supported for multi-page tables
\usepackage{array}      % Better column control
\usepackage{enumitem}
\usepackage{microtype}
\AtBeginDocument{%
  }

\copyrightyear{2026}
\acmYear{2026}
\setcopyright{cc}
\setcctype{by-nc-nd}
\acmConference[CHI EA '26]{Extended Abstracts of the 2026 CHI Conference on Human Factors in Computing Systems}{April 13--17, 2026}{Barcelona, Spain}
\acmBooktitle{Extended Abstracts of the 2026 CHI Conference on Human Factors in Computing Systems (CHI EA '26), April 13--17, 2026, Barcelona, Spain}
\acmDOI{10.1145/3772363.3799294}
\acmISBN{979-8-4007-2281-3/2026/04}

\begin{document}

\title{Topic Matters: How Linguistic Properties can Shape Reading Behaviour in Selective Exposure Studies}

\author{Thomas Krämer}
%\authornote{}
\email{thomas.kraemer@gesis.org}
\orcid{0000-0003-0507-7843}
\affiliation{
  \institution{GESIS - Leibniz Institute for the Social Sciences}
  \city{Cologne}
  \country{Germany}
}

\author{Dagmar Kern}
\email{dagmar.kern@gesis.org}
\orcid{0000-0003-1794-625X}
\affiliation{
  \institution{GESIS - Leibniz Institute for the Social Sciences}
  \city{Cologne}
  \country{Germany}
}

\author{Thomas Kosch}
\email{thomas.kosch@hu-berlin.de}
\orcid{0000-0001-6300-9035}
\affiliation{
  \institution{HU Berlin}
  \city{Berlin}
  \country{Germany}
}

\author{Daniel Hienert}
\email{daniel.hienert@gesis.org}
\orcid{0000-0002-2388-4609}
\affiliation{
  \institution{GESIS - Leibniz Institute for the Social Sciences}
  \city{Cologne}
  \country{Germany}
}  
\renewcommand{\shortauthors}{Krämer et al.}

\begin{abstract}
Research on selective exposure frequently relies on eye tracking to study reading behaviour, often assuming that texts across different controversial topics are comparable once basic controls are applied. This assumption is problematic if topic-dependent linguistic properties systematically shape how users read and allocate attention. We therefore examine whether such properties relate to differences in reading behaviour in selective exposure contexts. We analyse linguistic features and eye-tracking data from a laboratory study in which 68 participants searched for and read news articles on climate change and migration policy. Our results reveal systematic differences in both textual characteristics and reading behaviour across topics. These findings identify an important methodological confound in selective exposure research and highlight the need to account for topic-specific linguistic properties when interpreting eye-tracking measures and designing systems intended to mitigate biased information consumption. 
\end{abstract}



\maketitle

\section{Introduction}\label{introduction-related-work}

Public discourse in online information environments is increasingly shaped by controversial and polarizing topics. As users are confronted with large volumes of potentially conflicting information, concerns have grown about how individuals select and attend to information. One prominent explanation for these patterns of information consumption is selective exposure~\cite{sears1967selective}, understood as a behavioural manifestation of the broader phenomenon of confirmation bias~\cite{nickerson1998confirmation}, both of which have long been central topics in political science and are discussed as key mechanisms contributing to echo chambers and filter bubbles (e.g., \cite{moller2021filter, Garrett2009a, Knobloch-Westerwick2011}). In recent years, these concepts have also garnered increasing attention in Human–Computer Interaction (HCI), particularly in the form of cognitive biases \cite{Boonprakong2025a}. HCI research extends this discourse by investigating how interface design and system features can either exacerbate or mitigate selective exposure of information~\cite{kraemer2025chiCostelloWorkshop}.

A substantial body of HCI work explores design interventions aimed at reducing selective exposure, for example, through source position indicators \cite{Liao2014}, source expertise indicators \cite{Liao2014a}, stance and credibility labels \cite{Gao2018}, or more broadly through the development of bias-aware systems \cite{BoonprakongDinglerTag2023}. A fundamental prerequisite for such systems is the reliable identification of selective or biased user behaviour. To this end, researchers employ a variety of behavioural and physiological measures, including interaction logs  \cite{Garrett2009, Knobloch-Westerwick2011}, eye-tracking data \cite{Suelflow2019, ZillichGuenther2021}, and other physiological signals such as electrodermal activity, and brain hemodynamic responses \cite{BoonprakongDinglerTag2023}, as well as electroencephalography (EEG)~\cite{Kraemer2025}.

Recent advances in eye-tracking technology, such as the availability of webcam-based approaches and growing possibilities for real-time analysis using Artificial Intelligence~\footnote{\url{https://sdk.eyedid.ai/}}, or web-based use of professional eye-trackers outside the lab \cite{RoehrlEtAl2025}, lead to an increased interest in detecting selective exposure through eye-movement behaviour. Eye movements provide indicators of cognitive processing during reading. For instance, longer fixation durations are commonly associated with higher processing load \cite{just1980theory}. Fixations, saccades, and derived reading measures vary with individual differences \cite{shi2024effects}, as well as with lexical properties and comprehension demands \cite{reichle2006z, jarodzka2017tracking, rayner1998eye}. 
%Importantly, these factors are strongly influenced by the textual style in which information is presented.

Empirical studies on selective exposure, with and without eye tracking, frequently rely on controlled experimental settings, often using mock-up websites to simulate news feeds (e.g., Twitter/X in \cite{Boonprakong2025b} or Facebook in \cite{Suelflow2019}), partisan online news sites \cite{Garrett2009}, or curated sets of arguments that participants are asked to read and evaluate \cite{Liao2014}. The primary advantage of these mock-ups lies in the researchers’ ability to control textual properties, such as text length \cite{Boonprakong2025b, Liao2013} or readability (e.g., Flesch Reading Ease scores; \cite{BoonprakongDinglerTag2023}). Some studies additionally annotate texts with respect to stance \cite{Draws2022} or evaluative polarity (positive, negative, neutral, as in \cite{ZillichGuenther2021}). Alongside these controlled approaches, more naturalistic studies have examined selective exposure in free web searches \cite{ZillichGuenther2021} or news search tasks \cite{Kraemer2025}, without the ability to control the selected and read text.

Across both controlled and naturalistic studies, researchers typically investigate selective exposure and confirmation bias in the context of controversial or highly debated topics, including abortion rights \cite{Boonprakong2025b}, political progressivism \cite{BoonprakongDinglerTag2023}, the death penalty \cite{Liao2014, Liao2014a}, vegetarianism \cite{Boonprakong2025b, Liao2014}, same-sex marriage \cite{Garrett2009}, migration \cite{Kraemer2025, Suelflow2019}, and climate change \cite{Kraemer2025, BoonprakongDinglerTag2023}. While different topics are often used to test the robustness or generalizability of selective exposure effects, the texts associated with these topics are typically treated as functionally equivalent once basic controls (e.g., length or readability) are applied. 
At the same time, prior research has demonstrated that individual characteristics such as topic interest \cite{BoonprakongDinglerTag2023, Suelflow2019}, topic involvement \cite{Liao2013}, low-effortful thinking \cite{Boonprakong2025b}, and strong political beliefs \cite{Boonprakong2025b} shape selective exposure behaviour. However, comparatively little attention has been paid to how topic-specific textual properties themselves might influence selective exposure factors. 

In this paper, we analyse textual properties and reading behaviour from a user study on selective exposure using two controversial topics: climate change and migration policy. We observe systematic differences in both linguistic features and eye-movement patterns across the two topics.

\section{User Study and Dataset}\label{study}
We analysed data from a laboratory user study conducted in the context of selective exposure to address the following research question: 
How do linguistic properties of texts on different controversial topics relate to differences in reading behaviour in selective exposure contexts?

In the following, we describe the aspects of the user study relevant to answering this research question. 
%\subsection{Study Design and Procedure}
The user study included 68 participants (native German speakers, 42 female, 24 male, and 2 others, age: M=27, SD=7.9) and was designed to simulate a real-world scenario of searching for news online. Each participant completed two eight-minute search tasks~\footnote{Task: Your train is delayed, and you now have some unexpected free time. You decide to catch up on the current debate about [migration policy/climate change]. Use the news search and enter any search terms you like.} in a counterbalanced order.
%Participants used a self- implemented search engine employing BM25 as its ranking algorithm. 
The search engine~\footnote{Based on \url{https://www.elastic.co/elasticsearch}} indexed a corpus of 192,685 German-language news articles crawled from mainstream outlets via the Fundus news crawler~\footnote{  \hyperlink{https://github.com/flairNLP/fundus/tree/master/src/fundus/publishers/de}{https://github.com/flairNLP/fundus/tree/master/src/fundus/publishers/de.}}. Participants completed one task on migration policy and one task on climate change. These topics were selected because they frequently trigger public controversy and polarized discourse \cite{Mau2023}. After the search and reading phase, participants rated a subset of sentences for familiarity and agreement. These data are not analysed in the present paper. 
The study protocol received institutional ethics approval.

%\subsection{Eye-Tracking Data Collection and Processing}
The study was conducted in a laboratory setting, and eye movements were recorded using a Tobii Pro Spectrum eye tracker at 300\,Hz. We used EyeLiveMetrics \cite{Hienert2024} to compute eye-tracking metrics on the sentences read by participants, including fixation metrics (e.g., Fixation Duration, Fixation Count), saccade metrics (e.g., Saccades Count, Rightward Saccades Length), and reading metrics (e.g., Regression-Path 
Duration, Re-Reading Duration). As EyeLiveMetrics provides word-level metrics, we aggregated these measures to the sentence level. To account for variation in sentence length, all metrics (except Rightward Saccades Average Length, Rightward Saccades Average Amplitude and Rightward Saccades Scan Speed) were normalized by the number of words.
%\subsection{Data set Construction}
To analyse the relationship between linguistic properties and reading behaviour, we restricted our analysis to the sentences participants read. We further applied the following filtering steps: Only the first visit to a page was included. Subsequent visits after navigating away were excluded. Sentences were included only if their total fixation duration was at least 100 ms, which is considered the minimum time required for basic lexical or semantic processing of one or two words of a sentence \cite{just1980theory, reichle2006z}. % 10_5_1  / 10_5_10
After filtering, our dataset included 10,030 sentences for migration policy and 10,387 for climate change. Sentences could appear multiple times in the dataset when read by different participants.

%\subsection{Linguistic Feature Extraction}
For analysing the textual properties, we used the ELFEN (Efficient Linguistic Feature Extraction for Natural Language Datasets) framework~\footnote{Details see \url{https://elfen.readthedocs.io/en/latest/tutorials.html\#ratio-normalize}}\cite{maurer2025ELFEN}. ELFEN has been developed and used for the classification of human-generated vs. LLM-generated content\cite{donmez-etal-2025-ai}. To enable an unbiased comparison of linguistic properties across topics, we considered only unique sentences for this analysis. This resulted in 4,095 sentences for migration policy (average sentence length 16.9 tokens, $SD=9.6$) and 4,381 unique sentences for climate change (average sentence length 17.3 tokens, $SD=10.4$)~\footnote{Tokenisation via spaCy with the German News transformer model \protect\url{https://spacy.io/models/de\#de_dep_news_trf}}. The average sentence length did not differ significantly between topics.
% 10_5_10

\section{Results and Discussion}\label{results}
\subsection{Effects of Topic on Stimulus Properties}
\label{subsec:stim_prop}
% 10_5_10
In the text-centric analysis, we analysed 208 different linguistic metrics for all read sentences computed by the ELFEN framework. The metrics are based on the entire sentence text, not only the fixated parts. The metrics were normalized by the number of tokens (e.g., words, punctuation, or white space) in a sentence.
%Dimension reduction
Some metrics are expected to be highly correlated, for example, average word length (avg\_word\_length) and the number of long words (n\_long\_words). We therefore first computed a Spearman's rank correlation matrix, then clustered metrics with a correlation threshold of $>0.8$, and selected one representative metric per cluster with the highest variance, reducing the number to 148 metrics and limiting the over-representation of closely related metrics.
%In future analysis, we will consider different approaches for dimension reduction, such as manual, theory-driven selection of metrics, or algorithmic alternatives such as Principal Component Analysis (PCA) or regularized regression (e.g., LASSO). 
% LM!
We computed a Linear Model~\footnote{R package lm()} for each z-scaled linguistic metric across topics, allowing the regression coefficients to be interpreted and compared as standardized effect sizes. 
%The results of all models were then corrected for multiple comparisons.
% Significance / FDR 
We control the False Discovery Rate (FDR) by applying Benjamini-Hochberg correction. 
We found a significant difference ($p < .001$) between the two topics, migration policy and climate change, for 26 linguistic metrics.

\begin{figure*}[h!]
    \centering
    \includegraphics[width=1\linewidth]{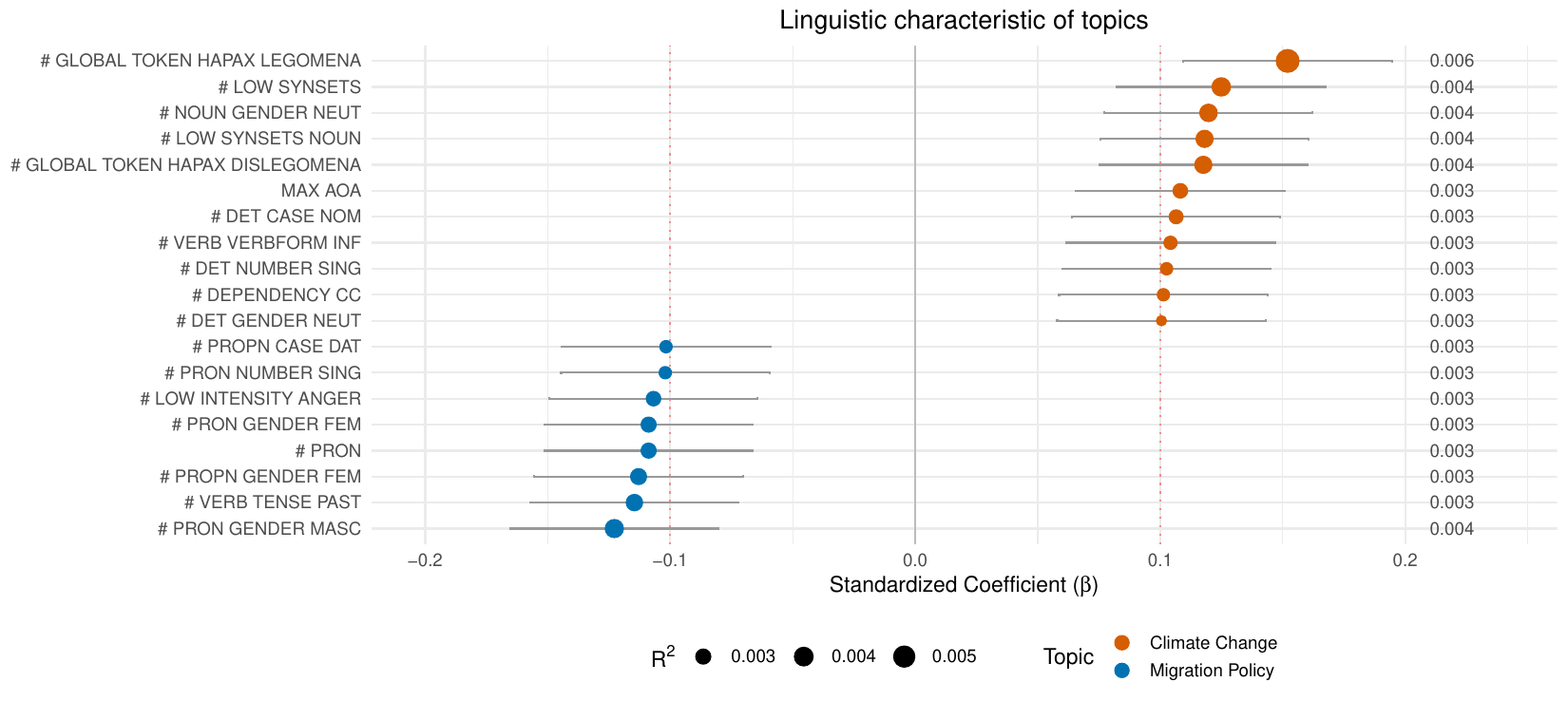}
    \caption{Linguistic metrics: effects with at least 0.1 standard deviation difference and maximum 0.001 $p_{adj}$ (FDR-corrected significance level). The circle sizes refer to $R^2$. With migration policy as reference category in the linear model, $\beta>0$ means the feature ratio is higher in Climate Change sentences, $\beta<0$ means the feature ratio is higher in Migration Policy sentences (in standardized units).}
    \Description{Plot summarizing topic differences in word-normalized linguistic feature ratios (ELFEN metrics) between Migration Policy and Climate Change sentences. Each row corresponds to one linguistic feature retained after redundancy reduction based on Spearman rank correlations among features (hierarchical clustering of $| \rho | > .80$; one representative per cluster). Points show the standardized regression coefficient $\beta$ from a linear model with the z-scored feature value as dependent variable and topic as predictor ($\beta = \text{Climate Change} - \text{Migration Policy}$, in standard-deviation units). Horizontal error bars indicate 95\% confidence intervals ($\beta \pm 1.96\,SE$). The solid vertical line at $\beta=0$ marks no topic difference; dotted red lines at $\beta=\pm 0.1$ indicate the minimum effect-size threshold used for display. Positive coefficients indicate higher feature values in Climate Change, negative coefficients indicate higher values in Migration Policy. Point colour encodes the topic with the higher mean feature value (orange: Climate Change; blue: Migration Policy), point size encodes $R^2$ (variance in the feature explained by topic), and the numeric labels on the right report the corresponding $R^2$. Only features with BH/FDR-adjusted $p<.001$ and $|\beta|\ge 0.1$ are shown.}
    \label{fig:ling_forest_plot}
\end{figure*}

% Practical Significance / Effect Magnitude
In Figure \ref{fig:ling_forest_plot}, only those metrics are visualized that reached a minimum standardized effect size of $|\beta| \ge 0.1$ to filter out negligible variations that have $p_{adj} \leq 0.001$. The labels refer to the acronyms and definitions from Universal Dependencies~\footnote{\url{https://universaldependencies.org/docs/}} as also referred to in the ELFEN Framework. The analysis reveals linguistic divergence across 19 metrics between the two topics. The circle size represents the variance explained by the topic ($R^2$), and the position indicates the magnitude of the difference. 

The metrics with the highest effect measured for climate change are the number of hapax legomena (\# GLOBAL TOKEN HAPAX LEGOMENA, $\beta=+0.15$), indicating tokens that occur only once in our sentence corpus, and the number of words with few synonyms (\# LOW SYNSETS, $\beta=+0.12$), indicating more specialized vocabulary. For migration policy, the metrics with the highest effects are the the number of verbs in past tense (\# VERB TENSE PAST, $\beta=-0.12$) and pronouns with male genus (\# PRON GENDER MASC, $\beta=-0.12$).  

While the topics are statistically distinguishable across many linguistic dimensions (26 metrics), the low $R^2$ values suggest that topic-specific vocabulary and grammar reflect minor differences, and hence the results should be interpreted with caution. For instance, 
climate change exhibited a slightly more technical and informational profile, defined by high noun density, a diverse and specialized vocabulary (hapax legomena), and higher Age of Acquisition (AOA) scores, while migration policy appears as a more narrative style, utilizing significantly more personal pronouns and past-tense verbs, indicating a focus on human actors and historical events. 

\subsection{Effects of Topic on Reading Behaviour}
\label{subsec:gaze}
% 10_5_2
Except Rightward Saccades Average Length, Rightward Saccades Average Amplitude and Rightward Saccades Scan Speed, eye-tracking measures were word-normalized to capture visual processing effort per unit of text rather than total sentence length, thereby aligning the gaze analysis with the sentence-level linguistic analysis. 
We modelled sentence-level gaze metrics using Generalized Linear Mixed Models (GLMM) to account for the hierarchical data structure (sentences nested within participants) and unbalanced trial numbers. For each gaze metric, we fit a model with the factor topic, with migration policy as reference category, and random intercepts for participant and sentence.

\begin{table}[htp]
\caption{\label{tab:tab:eye_tracking_results}Differences between eye-tracking metrics with respect to the reference category migration policy to climate change in percent. Negative values correspond to lower values for sentences read in the topic climate change compared to migration policy. All metrics are normalized by word per sentence, if not otherwise stated.}
\Description{Table showing percentage differences in eye-tracking metrics 
between climate change and migration policy conditions. 
Largest significant reductions are observed for First-Pass Regression-Path Duration(-11.9\%), 
First-Pass Re-Reading Duration (-13.6\%), and First-Pass Selective Regression-Path Duration(-9.6\%). 
Most other metrics show small and non-significant differences.}
\centering
\fontsize{7pt}{9pt}\selectfont % 9pt font with 11pt line spacing
\centering
\begin{tabular}[t]{lrrrrlr}
\toprule
Metric & $\%\Delta$ & $p$& Sig & SE &  $R^2_{m}$&$R^2_{c}$\\
\midrule
1st Pass Regression-Path Duration& -11.9 & <.001 & *** & .034 &  .002&.407 \\
1st Pass Re-Reading Duration & -13.6 & <.01 & ** & .048 &  .002&.396 \\
1st Pass SRPD  & -9.6 & <.01 & ** & .034 &  .001&.429 \\
 Look Back Time (>= 2nd Pass) & -4.5 & .034 & * & .022 &  .000&.440 \\
 1st Pass First Fixation Duration & -4.8 & .043 & * & .024 &  .000&.543 \\
\addlinespace
 Fixation Duration & -3.2 & .077 & n.s. & .019 &  .000&.482 \\
 Fixation Count & -3.1 & .083 & n.s. & .018 &  .000&.485 \\
 Saccades Count & 3.4 & .263 & n.s. & .030 &  .000&.456 \\
 Rightward Saccades ($\varnothing$ amplitude) & -2.3 & .278 & n.s. & .021 &  .000&.396 \\
 Rightward Saccades Count & 3.1 & .312 & n.s. & .030 &  .000&.453 \\
 Rightward Scan Speed & 1.1 & .391 & n.s. & .013 &  .000&.294 \\
 1st Pass Duration & -1.1 & .633 & n.s. & .023 &  .000&.398 \\
 1st Pass Regression & -0.7 & .719 & n.s. & .021 &  .000&.192 \\
 Rightward Saccades Length & -.6 & .854 & n.s. & .031 &  .000&.154 \\
 Rightward Saccades Length ($\varnothing$ in px) & 0.0 & .999 & n.s. & .021 &  .000&.373 \\
\bottomrule
\end{tabular}
\end{table}

Table ~\ref{tab:tab:eye_tracking_results} shows differences of gaze metrics between the two topics. For the First-Pass Fixation Duration the -4.8\% difference ($p = .043$) suggests that migration-related vocabulary required slightly more effort during early lexical processing. Regression metrics in the first-pass reading of a sentence, such as Re-Reading Duration, Regression Path Duration, and Selective Regression Path Duration, showed the most substantial effects, with higher durations for migration policy (9.6\% to 13.6\% increase, $p<.01$). This suggests that participants return significantly longer to previous parts of the sentence during the first-pass reading of the topic migration policy. Higher Regression-Path Duration indicates difficulties in integrating words when they are fixated ~\cite[p.~349]{clifton2007eye}, while Re-Reading Duration indicates general comprehension difficulties ~\cite[p.~363]{clifton2007eye} or provides information about revision and reanalysis ~\cite{CLIFTON20161}. For Look-back Time, the 4.5\% change ($p = .034$) for the topic migration policy indicates that readers were significantly more likely to return to previous sentences after reading further in the text to resolve, for example, ambiguities. 

\subsection{Relationship of Reading Behaviour and Linguistic Sentence Properties}
% 10_5_11
To better understand the relationship between linguistic properties of the text and gaze behaviour, we computed the correlation matrix of the 19 linguistic metrics identified in Section \ref{subsec:stim_prop} and the 5 significant gaze metrics from Section \ref{subsec:gaze}. 
Due to the characteristic right-skew of gaze metrics (confirmed via Q-Q plots and Shapiro-Wilk tests), Spearman’s rank correlations were employed to provide a robust measure of the relationship between linguistic features and gaze metrics 

\begin{figure*}[h!]
    \centering
    \includegraphics[width=1\linewidth]{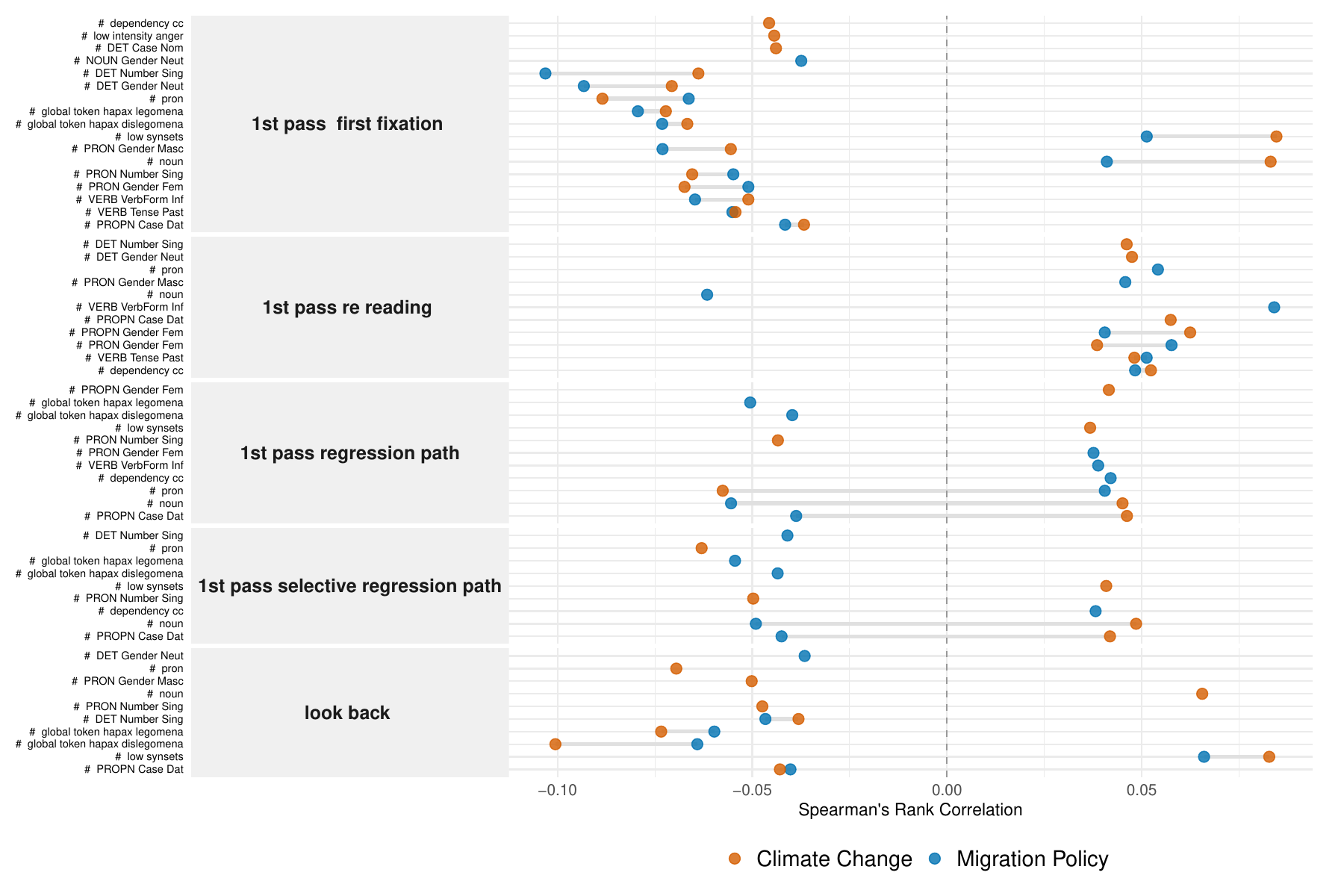}
    \caption{Linguistic metrics with influence on eye-movement behaviour. Features shown are those that reached statistical significance (Benjamini-Hochberg-corrected $p \leq .001$) in either or both the Migration Policy and Climate Change sentences.}
    \label{fig:sig_drivers_by_metric}
    \Description{Faceted plot of Spearman’s rank correlations ($\rho$) between word-normalized linguistic feature ratios (y-axis; e.g., proportion of determiners, pronouns, hapax terms) and sentence-level gaze metrics (one panel per gaze metric) computed separately for Migration Policy (blue) and Climate Change (orange). For each topic, Spearman’s $\rho$ is computed by ranking sentences on the linguistic feature and on the corresponding gaze metric and correlating these ranks (monotonic association; $\rho \in [-1,1]$). Points indicate topic-specific $\rho$ values; grey connecting segments appear when a feature-metric pair is significant in both topics, and their length reflects the topic difference in $\rho$. Positive values (right of zero) indicate that higher feature density is associated with larger gaze-metric values (e.g., longer durations/more time), whereas negative values indicate the opposite. The dashed vertical line marks $\rho=0$ (no monotonic association). The range ($\rho \in [-0.1,0.1]$) reflects small effect sizes: correlations near $\pm 0.05$ indicate very weak monotonic associations, and values approaching $\pm 0.10$ indicate weak associations, despite statistical significance due to sample size and FDR correction.}
    
\end{figure*}
Again, given the high dimensionality of the data, we controlled the False Discovery Rate (FDR) while maintaining sufficient statistical power to detect meaningful associations that might be suppressed by more conservative family-wise error rate corrections (e.g., Bonferroni). 
In Figure \ref{fig:sig_drivers_by_metric} only those associations are shown that remained significant after Benjamini-Hochberg correction, with a significance threshold set to $\alpha = 0.001$. 

We observed several correlations between linguistic and gaze metrics. % 1st pass FD, same for both topics 
First-Pass First Fixation Duration is affected similarly across both topics: Determiners and pronouns (\# DET Number Sing, \# DET Gender Neut, and \# PRON Gender Masc/Fem) show the strongest negative correlations ($\rho \approx -0.05$ to $-0.10$), indicating that higher proportions of these features are associated with shorter first-fixation durations on a word. A similar pattern is observed for infinitives and past tense verbs. First-Pass First Fixation Duration is surprisingly also reduced with a higher density of words that occur only once (Global Hapax Legomena) or twice (Dislegomena) in the sentence corpus. Because Hapax Legomena are defined relative to our specific stimulus corpus rather than a general language corpus, even words that are frequent in the language can function as hapaxes if they occur only once in our dataset. Only a higher density of nouns and of words with few synonyms consistently increased First-Pass First Fixation Duration for both topics. 

%(s)rpd, sign flip for --M/++C, except: srdp for pronouns ++M/--C
Our results show topic-dependent associations between \newline Regression-Path Duration and Selective Regression-Path Duration and noun-related features. Noun density and dative proper-noun density have a negative correlation in migration policy but a positive one in climate change. Pronoun density reverses sign (positive in migration policy and negative in climate change) for Regression-Path Duration. One possible explanation is that in migration discourse, nouns might refer to concrete actors, whereas in climate change, nouns might be more abstract, technical concepts that are harder to capture.

% Re-reading
The positive correlations between pronouns and verbs and Re-Reading Duration suggest that these features increase late-stage integration effort. Although identified quickly during the First-Pass First Fixation, they appear more difficult to incorporate into the broader sentence structure.
% Look back time, same direction for both topics
Look-back Time patterns are consistent across both topics: Low synonym density increases Look-back time~\footnote{thresholds for low synonym density: 2 or less synonyms}, while Hapax (Dis)Legomena reduces it.

\section{Conclusion}\label{conclusion}
In this paper, we analysed textual properties and reading behaviour from a selective exposure user study comparing news articles on two controversial topics: climate change and migration policy. By examining linguistic characteristics of the texts alongside participants’ eye-movement patterns, we identified systematic differences in both textual features and reading behaviour across topics. While the observed correlations between textual features and reading measures are relatively small, their statistical significance across thousands of observations suggests that linguistic features reliably capture aspects of processing effort during natural reading. In the context of the complex cognitive processes involved in reading, even modest effects can account for consistent variance in behavioural measures.

Taken together, these findings indicate that topic-dependent textual structure can meaningfully influence reading behaviour and, consequently, the interpretation of selective exposure measures based on eye-tracking data. Our results highlight the importance of accounting for topic-specific textual characteristics in selective exposure research and in the design of systems intended to mitigate biased information consumption, rather than assuming equivalence across content domains.
Furthermore, in the context of bias-aware systems, user interfaces might take these findings into account and dynamically adapt, e.g., via glossary overlays for technical content to support specialized vocabulary, or with context anchors, such as pronoun-referent highlighting in narrative content. 
\begin{acks}
For this publication, AI was used in accordance with the AI guidelines of our institution and the ACM. During the preparation of this work, ChatGPT v5 and Gemini v3 were used for grammar and style editing and for minor coding tasks. All content was reviewed and edited by the authors, who take full responsibility for the accuracy and final publication of this work.

This work is funded by the German Research Foundation (DFG) project ``Overcome Selective Exposure in Web Search by Considering Eye Movements and Physiological Signals''~(Project-ID~525041402). 
Furthermore, this work is supported by the German Research Foundation (DFG), CRC 1404: ``FONDA: Foundations of Workflows for Large-Scale Scientific Data Analysis''~(Project-ID~414984028).
\end{acks}

\bibliographystyle{ACM-Reference-Format}
\bibliography{bib}

\end{document}